\documentclass[aps,prc,twocolumn,superscriptaddress,nofootinbib,longbibliography,floatfix,10pt]{revtex4-2}

\usepackage{graphicx}
\usepackage{dcolumn}
\usepackage{bm}
\usepackage{morefloats}
\usepackage{multirow}
\usepackage{amssymb}
\usepackage{amsmath}
\usepackage{xcolor}
\usepackage{longtable}
\usepackage{fix-cm}
\usepackage{mathptmx} 
\usepackage[T1]{fontenc}

\usepackage[colorlinks,allcolors=blue]{hyperref}
\makeatletter
\def\NAT@def@citea{\def\@citea{\NAT@separator}}
\makeatother

\begin{document}

\title{Multinucleon transfer mechanism in ${}^{250}\text{Cf}+{}^{232}\text{Th}$
collisions using the quantal transport description based on the stochastic mean-field
approach}

\author{S. Ayik}\email{ayik@tntech.edu}
\affiliation{Physics Department, Tennessee Technological University, Cookeville, TN 38505, USA}
\author{M. Arik}
\affiliation{Physics Department, Middle East Technical University, 06800 Ankara, Turkey}
\author{O. Yilmaz}
\affiliation{Physics Department, Middle East Technical University, 06800 Ankara, Turkey}
\author{B. Yilmaz}
\affiliation{Physics Department, Faculty of Sciences, Ankara University, 06100 Ankara, Turkey}
\author{A. S. Umar}
\affiliation{Department of Physics and Astronomy, Vanderbilt University, Nashville, TN 37235, USA}

\date{\today}

\begin{abstract}
Production cross-sections of heavy neutron-rich isotopes are calculated by
employing quantal transport description in ${}^{250}\text{Cf}+{}^{232}\text{Th}$
collisions. This quantal transport description is based on the stochastic
mean-field (SMF) approach, and it provides a microscopic approach beyond time-dependent
Hartree-Fock (TDHF) theory to
include mean-field fluctuations. De-excitation of primary fragments is
determined by employing the statistical GEMINI++ code. Calculations provide
predictions for production cross-sections of neutron rich transfermium isotopes
without any adjustable parameters. 
\end{abstract}

\maketitle

\section{Introduction}
In recent years extensive experimental and theoretical investigations have been
carried out for the production of heavy elements close to the superheavy island
with proton numbers 
$Z>100$~\cite{chatillon2006,herzberg2001,kozulin2012,kratz2013,watanabe2015,desai2019,adamian2010,adamian2010b,jiang2020,adamian2020,kalandarov2020,li2020b,adamian2021,itkis2022,heinz2022,wu2022}. 
Fusion reactions are a natural mechanism for the
production of very heavy nuclei. Superheavy elements are identified by following the
decay pattern of the primary heavy fusion product. Thus far super heavy elements
have been synthesized either in cold fusion reactions~\cite{hofmann2000} or in hot fusion
reactions~\cite{oganessian2006,itkis2022}, using actinide nuclei. 
Highly excited compound nuclei de-excite
mostly by neutron emission and secondary fission. As a result, fusion reactions
may not be the most efficient way to produce neutron rich heavy isotopes. 

As an alternative mechanism, for the production of neutron rich heavy isotopes,
multi-nucleon transfer (MNT) processes have been experimentally investigated with
actinide targets near barrier energies, and more investigation are currently in
progress. Such investigations may provide a more efficient mechanism for the production
of heavy neutron-rich isotopes. Multinucleon transfer mechanism has been
investigated using several phenomenological approaches including
multidimensional Langevin 
model~\cite{zagrebaev2006,zagrebaev2008,zagrebaev2008c,zagrebaev2011,karpov2017,saiko2019,saiko2022}, 
di-nuclear system (DNS) model~\cite{feng2009,feng2009a,feng2017} and
quantum molecular dynamics (QMD) model~\cite{zhao2009,zhao2016,wang2016}. In order provide more accurate
description of collision dynamics and for MNT mechanism it is
important to develop microscopic approaches, which also provides a test for phenomenological models. 
Time-dependent Hartree-Fock (TDHF)
theory provides a microscopic description for the mean evolution of collective
dynamics at low bombarding 
energies~\cite{simenel2012,simenel2018,nakatsukasa2016,oberacker2014,umar2015a,umar2015c,umar2017,simenel2010,sekizawa2016,sekizawa2019}. 
However, the TDHF theory has a
severe limitation: it can only describe the most probable dynamical path of the
collision dynamics with small fluctuations around it. It describes mean kinetic
energy loss due to one body dissipation rather well, but it cannot describe the
large dispersions of mass and charge distribution of the fragments. 
To remedy this problem one must go beyond TDHF~\cite{tohyama2002a,tohyama2020,simenel2011,lacroix2014}.
The time-dependent random phase approximation (TDRPA) of Balian and V\'en\'eroni provides
an important improvement of the mean-field description. This approach has been
applied for analysis of multinucleon transfer in several 
studies~\cite{balian1985,balian1992,williams2018,godbey2020}.
However, the approach is limited to calculate dispersions of charge and mass
distributions in symmetric collisions. 

The stochastic mean-field (SMF) approach provides a further improvement of the
TDHF theory beyond the mean field approximation~\cite{ayik2008,lacroix2014}, and can be
applied to asymmetric collisions. 
In Sec.~\ref{sec2} we present
results of TDHF calculations for the collisions of the
${}^{250}\text{Cf}+{}^{232}\text{Th}$ system at $E_{\text{c.m.}}=950$~MeV. In
Sec.~\ref{sec3}, we briefly describe the quantal transport description of
multi-nucleon transfer based on the SMF approach. We present an analysis of
multinucleon transfer mechanism for the same reaction. This analysis is essentially
a complimentary description of the work of Kedziora and Simenel~\cite{kedziora2010}, in which
multinucleon transfer mechanism has been investigated in the TDHF approximation alone.
In Sec.~\ref{sec4}, we present results of quantal transport description based on SMF
approach for ${}^{250}\text{Cf}+{}^{232}\text{Th}$ reaction. Our calculations
can describe primary and secondary isotope production cross-sections including
mean values and fluctuations without any adjustable parameters except standard
parameters of the Skyrme energy density functional. In Sec.~\ref{sec5}, conclusions are
given.

\section{Mean-Field Description}
\label{sec2}
The microscopic TDHF theory, employing
effective Skyrme type energy density functionals, has been used extensively for
describing heavy-ion collisions and nuclear 
fusion~\cite{simenel2012,nakatsukasa2016,oberacker2014,umar2015a,simenel2010,sekizawa2016,simenel2018,sekizawa2019}. 
The mean-field
theory provides a good description for the most probable dynamical path of the
collective motion at low energy heavy-ion collisions, including the one-body
dissipation mechanism. The TDHF is a deterministic approach for many-body
dynamics in the sense that mean-field evolution starting with a given initial
condition leads to a single, deterministic final state. For example, for a
collision with a given charge and mass asymmetry it leads to a single exit channel with a
certain charge and mass asymmetry. 

In ${}^{250}\text{Cf}+{}^{232}\text{Th}$
system, both projectile and target nuclei exhibit strong prolate deformation in
their ground states. As a result, the collision dynamics and multinucleon
transfer mechanism, strongly depend on the collision geometry. We consider four
different initial collision geometries at the same bombarding energy
$E_{\text{c.m.}}=950$~MeV. In analogy with work of Kedziora and Simenel
of Ref.~\cite{kedziora2010},
we indicate initial orientation of target or projectile along the beam
direction with letter  $X$, and perpendicular to the beam direction
with letter $Y$. Four different collision geometries are represented as
$\text{XX}$, $\text{XY}$, $\text{YX}$, $\text{YY}$ and correspond to tip-tip,
tip-side, side-tip and side-side geometries, respectively (see Fig.~2 and
Fig.~3 of Ref.~\cite{kedziora2010}). 
As a convention, the first letter indicates the initially heavy
partner of the colliding system. Table~\ref{tab1} exhibits results of TDHF calculations
for different values of initial orbital angular momentum $\ell_i$, final values of
mass and charge numbers of $\text{Cf}$-like $A_1^f$, $Z_1^f$ and
$\text{Th}$-like $A_2^f$, $Z_2^f$ fragments, final total kinetic energy
($\text{TKE}$), total excitation energy ($E^{\ast}$), scattering angles in the
center of mass frame $\theta_{\text{c.m.}}$ and lab frame $\theta_{1}^{lab}$ and
$\theta_{2}^{lab}$ for the four different collision geometries. The table also
includes asymptotic values of neutron $\sigma_{\text{NN}}$, proton
$\sigma_{\text{ZZ}}$, mixed dispersions $\sigma_{\text{NZ}}$, and mass dispersion
$\sigma_{\text{AA}}$, which will be discussed in Sec.~\ref{sec4}(A). To reduce
computation time, we have chosen the initial orbital angular momentum in steps
of 40$\hbar$.
The calculations presented in the rest of the article employed the TDHF 
code~\cite{umar1991a,umar2006c} using the SLy4d Skyrme energy density functional~\cite{kim1997},
with a box size of $60\times 60\times 36$~fm in the $x-y-z$ directions, respectively.

As seen from Table~\ref{tab1}, distinct
geometries result in different nucleon transfer mechanism. The
different nucleon transfer mechanisms for head-on collisions is observed more clearly
from the time evolution of neutron $N(t)$ and proton  $Z(t)$ numbers of
$\text{Cf}$-like fragments or $\text{Th}$-like fragments. In Fig.~\ref{fig1A} of 
Appendix~\ref{appA}, we plot
the time evolution of neutron and proton numbers for the $\text{Cf}$-like fragments
for the four different geometries studied. In
standard TDHF calculations, time evolution of macroscopic variables for a di-nuclear
complex, such as the charge and mass of target-like or projectile-like fragments, are not
utilized explicitly. Often just the values of macroscopic variables at exit
channel are employed in the analysis of the reaction mechanism. Drift paths, which
show the evolution of the system in the $N$-$Z$ plane, carry a more detailed
information about the nucleon transfer mechanisms. As a result of shell effects on
dynamics, drift paths exhibit different behavior in different collision
geometries and include detailed information of time evolution of the mean values
of macroscopic variables. On the other hand, the time evolution of macroscopic variables
becomes very
important for the diffusion mechanism, as we discuss in Sec.~\ref{sec4}. The blue curves in
Fig.~\ref{fig1} show drift paths in head-on collisions of different geometries
for Cf-like fragments. In these figures, thick black lines indicate equilibrium
charge asymmetry with $(N-Z)/(N+Z)=0.22$. This line is referred to as the
isoscalar path, which follows nearly parallel to the bottom of the stability line.
The isoscalar path extends all the way toward the lead valley on one end, and toward
the superheavy valley on the other end, making about $\phi =32^{\circ}$ angle
with respect to the horizontal neutron axis. We observe that in all geometries, 
$\text{Cf}$-like fragments drift nearly along the isoscalar direction with charge
asymmetry approximately equal to $0.22$. Figure~\ref{fig1}(a) shows the drift
path for the tip-tip collision. As usually observed in quasifission reactions,
$\text{Cf}$-like heavy fragments loose nucleons and the system drifts toward
symmetry. In the side-tip collision, shown in Fig.~\ref{fig1}(c), nucleon drift
mechanism is very different than the tip-tip geometry. Here, the heavy fragment gains neutrons
and protons and the system drifts along the isoscalar path toward asymmetry. This kind
of drift path is not very common, and it is referred to as the inverse quasifission
reaction. Figure~\ref{fig1}(d) shows the drift path for the side-side collision. As a
result of strong shell effects, the di-nuclear system appears to be at a near local
equilibrium state in the $N$-$Z$ plane. System initially drifts along the isoscalar path
toward symmetry. Subsequently, the drift stops and the di-nuclear system evolves along
the isoscalar path toward asymmetry and separates approximately with the entrance channel
charge and mass asymmetry. At the exit channel, it appears that system does not
exhibit any drift. As seen in Fig.~\ref{fig1}(b), in the tip-side collision, nucleon drift
mechanism is very different than those in other geometries. The di-nuclear
system drifts along isoscalar path with the same charge asymmetry toward
symmetry. However, $\text{Cf}$-like heavy fragments continue to lose neutrons
and protons until they nearly  reach thorium at the exit channel. 
In Ref.~\cite{kedziora2010},
this type of drift was named as swap inverse quasifission reaction. 

\section{Quantal Diffusion Description}
\label{sec3}
\subsection{Langevin equation for nucleon transfer}

The ordinary TDHF provides a deterministic description for collision dynamics. A
single-particle density matrix is calculated with a given initial condition
which is characterized by a single Slater determinant. On the other hand, due to
correlations the actual initial state cannot be a single determinant but should
be a superposition of Slater determinants. In the SMF approach, the correlated
initial sate is represented by an ensemble of single-particle density matrices
which are specified in terms of initial correlations~\cite{ayik2008,lacroix2014}. 
Time evolution of
single particle density matrix in each event in the ensemble is determined by
the TDHF equations with the self-consistent Hamiltonian of that event. In each
event of the SMF approach, fluctuations of the random elements of the initial
density matrices are determined by Gaussian distributions whose variances
are specified with the requirement that the ensemble average of dispersions of
one-body observables in the initial state match the quantal expressions
in the mean-field approach. 
\begin{table*}[!htb]
\caption{Results of the TDHF and SMF calculations for the
${}^{250}\text{Cf}+{}^{232}\text{Th}$ system at $E_\text{c.m.}=950$~MeV in tip-tip
(XX), tip-side (XY), side-tip (YX) and side-side (YY) geometries.}
\label{tab1}
\begin{ruledtabular}
\begin{tabular}{c | c c c c c c c c c c c c c c r }
- &$\ell_i$ $(\hbar)$ &$Z_1^f$ &$A_1^f$  &$Z_2^f$ & $A_2^f$ &$\ell_f$ $(\hbar)$ 
& $TKE$  &$E^*$     & $\sigma_{NN}$ & $\sigma_{ZZ}$ & $\sigma_{NZ}$ &
$\sigma_{AA}$  & $\theta_\text{c.m.}$ & $\theta_{1}^{lab}$ & $\theta_{2}^{lab}$  \\
\hline

\multirow{13}{2.5em}{\textbf{XX}} 
&0 &89.9 &230.7 &98.1 &251.3 &0.0 &612.8 &335.9 &8.6 &5.5 &5.5 &13.6 &180.0 &0.0 &0.0\\
&40 &90.7 &232.7 &97.3 &249.3 &27.1 &606.7 &342.6 &8.6 &5.6 &5.6 &13.7 &171.9 &3.7 &28.6\\
&80 &92.9 &238.4 &95.1 &243.6 &63.1 &603.9 &345.4 &8.7 &5.7 &5.7 &13.9 &163.2 &7.6 &47.6\\
&120 &95.9 &246.5 &92.1 &235.5 &92.4 &589.0 &362.6 &8.8 &5.7 &5.7 &14.0 &154.0 &11.5 &55.9\\
&160 &98.6 &253.3 &89.4 &228.7 &123.3 &571.1 &375.8 &8.9 &5.8 &5.8 &14.1 &144.5 &15.3 &57.8\\
&200 &98.9 &253.9 &89.1 &228.1 &147.9 &564.3 &381.4 &8.8 &5.7 &5.7 &14.1 &135.2 &19.1 &56.0\\
&240 &98.3 &252.7 &89.7 &229.3 &178.1 &563.6 &384.1 &8.8 &5.7 &5.7 &14.0 &125.3 &23.3 &53.1\\
&280 &96.9 &248.9 &91.1 &233.0 &219.5 &567.0 &382.3 &8.7 &5.6 &5.6 &13.8 &115.3 &27.8 &49.3\\
&320 &97.2 &249.7 &90.8 &232.3 &243.4 &590.4 &358.8 &8.4 &5.5 &5.5 &13.4 &106.9 &31.6 &47.1\\
&360 &97.5 &250.0 &90.5 &232.0 &258.7 &625.5 &324.5 &8.0 &5.2 &5.2 &12.7 &100.9 &34.7 &45.8\\
&400 &96.6 &247.5 &91.4 &234.4 &283.8 &668.3 &279.9 &7.4 &4.8 &4.8 &11.6 &95.8 &37.8 &44.4\\
&440 &96.7 &248.4 &91.3 &233.6 &330.9 &718.7 &229.5 &6.6 &4.3 &4.3 &10.3 &90.5 &41.0 &43.2\\
&480 &97.2 &249.1 &90.8 &232.9 &398.7 &774.9 &174.4 &5.6 &3.7 &3.7 &8.5 &85.7 &44.1 &42.0\\
\hline
\hline
\multirow{13}{2.5em}{\textbf{XY}} 
&0 &87.1 &222.4 &100.9 &259.6 &0.0 &627.4 &316.3 &7.8 &5.2 &5.8 &12.4 &180.0 &0.0 &0.0\\
&40 &87.5 &223.5 &100.5 &258.5 &30.7 &624.7 &322.7 &7.8 &5.2 &5.8 &12.4 &169.7 &4.9 &31.3\\
&80 &88.7 &227.0 &99.3 &255.0 &64.7 &625.0 &321.8 &7.8 &5.2 &5.8 &12.4 &159.1 &9.9 &47.7\\
&120 &90.3 &231.4 &97.7 &250.6 &98.6 &612.5 &336.2 &7.8 &5.2 &5.8 &12.5 &148.4 &14.6 &52.9\\
&160 &91.6 &235.1 &96.4 &246.9 &141.6 &592.2 &358.0 &7.9 &5.2 &5.8 &12.5 &137.1 &19.5 &52.7\\
&200 &93.2 &239.1 &94.8 &242.9 &181.7 &586.2 &364.0 &7.9 &5.2 &5.8 &12.5 &125.9 &24.2 &51.3\\
&240 &93.9 &241.3 &94.1 &240.7 &212.1 &591.6 &359.1 &7.8 &5.2 &5.8 &12.4 &116.4 &28.2 &49.1\\
&280 &95.4 &245.0 &92.6 &237.0 &244.7 &569.0 &380.9 &7.7 &5.1 &5.7 &12.2 &107.6 &31.3 &46.0\\
&320 &96.3 &247.4 &91.7 &234.6 &258.7 &607.0 &343.2 &7.4 &4.9 &5.4 &11.7 &102.1 &34.1 &45.4\\
&360 &96.8 &248.2 &91.2 &233.8 &286.5 &634.0 &314.2 &7.1 &4.8 &5.1 &11.2 &96.1 &37.0 &43.8\\
&400 &97.3 &249.6 &90.7 &232.4 &316.7 &667.6 &280.1 &6.8 &4.5 &4.8 &10.6 &91.4 &39.4 &42.7\\
&440 &97.3 &249.7 &90.7 &232.3 &363.7 &708.3 &239.4 &6.3 &4.2 &4.4 &9.8 &87.4 &42.0 &41.7\\
&480 &97.8 &251.3 &90.2 &230.7 &400.7 &745.3 &203.4 &5.7 &3.8 &3.8 &8.7 &83.5 &44.2 &40.7\\
\hline
\hline
\multirow{13}{2.5em}{\textbf{YX}} 
&0 &103.5 &266.5 &84.5 &215.5 &0.0 &647.7 &297.1 &7.6 &5.1 &5.4 &11.9 &180.0 &0.0 &0.0\\
&40 &103.6 &266.7 &84.4 &215.3 &30.8 &647.3 &296.4 &7.6 &5.1 &5.4 &11.9 &169.9 &4.4 &70.0\\
&80 &103.4 &266.3 &84.6 &215.7 &73.4 &644.5 &298.2 &7.6 &5.1 &5.4 &12.0 &159.6 &8.9 &71.7\\
&120 &103.2 &265.3 &84.8 &216.7 &96.8 &632.4 &311.6 &7.6 &5.1 &5.4 &12.0 &149.6 &13.1 &68.0\\
&160 &102.5 &263.4 &85.5 &218.6 &133.0 &613.8 &330.9 &7.6 &5.2 &5.5 &12.0 &139.1 &17.5 &62.8\\
&200 &101.4 &260.5 &86.6 &221.5 &177.4 &607.1 &336.6 &7.6 &5.2 &5.5 &12.0 &128.6 &22.0 &58.1\\
&240 &100.4 &257.3 &87.6 &224.7 &207.5 &601.3 &344.8 &7.6 &5.1 &5.5 &12.0 &119.0 &26.1 &53.5\\
&280 &99.0 &253.6 &89.0 &228.4 &238.9 &597.5 &348.2 &7.6 &5.1 &5.4 &11.9 &110.1 &30.1 &49.2\\
&320 &97.2 &249.9 &90.8 &232.1 &265.0 &615.7 &332.0 &7.4 &5.0 &5.2 &11.6 &103.1 &33.6 &46.4\\
&360 &96.5 &248.6 &91.5 &233.4 &286.2 &642.3 &307.0 &7.1 &4.8 &5.0 &11.2 &97.6 &36.4 &44.7\\
&400 &96.2 &247.5 &91.8 &234.5 &315.1 &668.2 &282.9 &6.8 &4.7 &4.7 &10.6 &92.3 &39.3 &42.8\\
&440 &96.1 &247.2 &91.9 &234.8 &346.4 &700.5 &249.7 &6.4 &4.4 &4.3 &9.9 &87.9 &41.9 &41.5\\
&480 &96.1 &247.2 &91.9 &234.8 &387.8 &740.5 &209.7 &5.9 &4.0 &3.8 &8.9 &84.2 &44.3 &40.5\\
\hline
\hline
\multirow{13}{2.5em}{\textbf{YY}} 
&0 &97.4 &250.0 &90.6 &232.0 &0.0 &620.0 &327.7 &9.3 &6.1 &6.8 &14.7 &180.0 &0.0 &0.0\\
&40 &98.2 &252.2 &89.8 &229.8 &27.3 &634.6 &315.2 &9.3 &6.1 &6.8 &14.7 &167.6 &5.5 &55.3\\
&80 &97.8 &251.5 &90.2 &230.5 &51.1 &632.9 &316.8 &9.2 &6.0 &6.8 &14.6 &155.0 &11.2 &62.1\\
&120 &97.5 &251.1 &90.5 &230.9 &90.9 &627.0 &321.7 &9.2 &6.0 &6.7 &14.5 &142.8 &16.5 &60.6\\
&160 &96.8 &248.9 &91.2 &233.1 &135.7 &613.6 &335.7 &9.1 &6.0 &6.7 &14.4 &130.7 &21.8 &56.2\\
&200 &96.7 &248.3 &91.3 &233.7 &177.0 &616.9 &331.3 &8.9 &5.8 &6.5 &14.1 &120.7 &26.3 &52.9\\
&240 &97.2 &249.7 &90.8 &232.3 &199.3 &639.5 &308.2 &8.6 &5.7 &6.2 &13.6 &113.1 &29.8 &51.2\\
&280 &97.0 &249.0 &91.0 &233.0 &236.8 &638.0 &311.3 &0.0 &5.5 &6.0 &13.2 &105.1 &33.2 &47.7\\
&320 &97.4 &249.8 &90.6 &232.2 &280.7 &627.6 &320.1 &8.2 &5.4 &5.8 &12.8 &97.5 &36.1 &44.4\\
&360 &97.6 &250.3 &90.4 &231.7 &312.6 &638.7 &310.6 &7.8 &5.1 &5.5 &12.1 &93.0 &38.1 &42.8\\
&400 &97.5 &250.4 &90.5 &231.6 &341.0 &659.1 &290.9 &7.3 &4.8 &5.1 &11.3 &87.5 &40.8 &40.8\\
&440 &97.5 &250.5 &90.5 &231.5 &378.4 &687.9 &260.8 &6.8 &4.5 &4.6 &10.5 &83.8 &42.9 &39.7\\
&480 &97.4 &250.3 &90.6 &231.7 &416.4 &725.6 &222.1 &6.3 &4.2 &4.0 &9.4 &81.2 &44.9 &39.2\\
\end{tabular}
\end{ruledtabular}
\end{table*}

When a di-nuclear structure is maintained in the collision dynamics, as in the
case of heavy-ion collisions at near barrier energies, we do not need to generate an
ensemble of mean-field events.
\begin{figure*}[!th]
\includegraphics*[width=15cm]{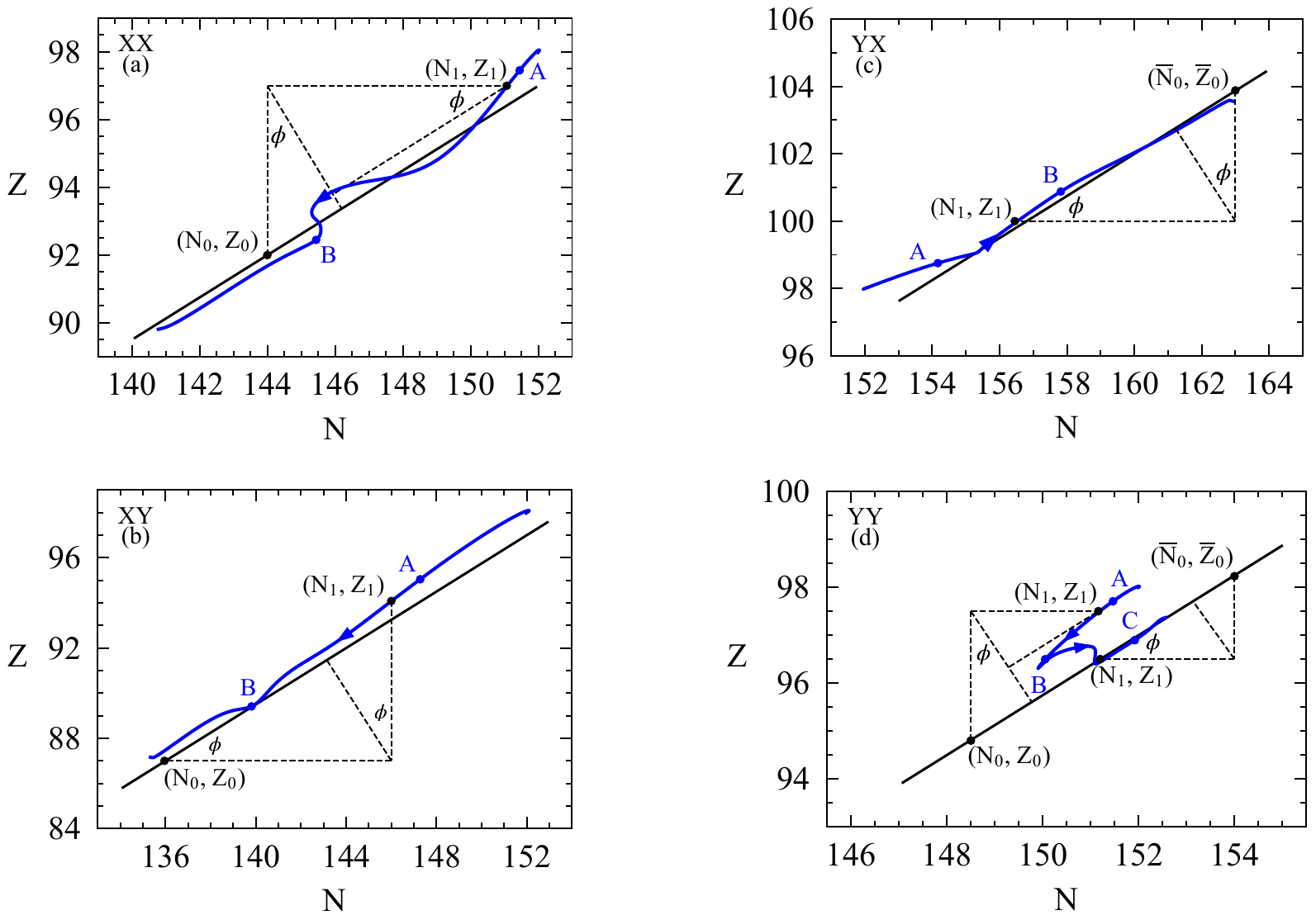}
\caption{Blue curves show drift path of Cf-like fragments in the head-on
collision of ${}^{250}\text{Cf}+{}^{232}\text{Th}$ system at $E_\text{c.m.}=950$~MeV 
in tip-tip (XX), tip-side (XY), side-tip (YX) and side-side (YY) geometries.
}
\label{fig1}
\end{figure*}
Instead, it is possible to develop a much easier transport description by
employing the Langevin formalism for the relevant macroscopic variables.
This is accomplished via a geometric
projection of the SMF approach by utilizing the window dynamics. For the details
of the quantal diffusion description and the window dynamics we refer to
Refs.~\cite{ayik2017,ayik2018,yilmaz2018,ayik2019,ayik2019b,sekizawa2020,ayik2020b,yilmaz2020,ayik2021}.
For the description of the nucleon diffusion mechanism, we consider
neutron number and proton number of the projectile-like or target-like fragments
as the relevant macroscopic variables. In this work, neutron $N_1^{\lambda
}(t)$, and proton $Z_1^{\lambda }(t)$, numbers of the $\text{Cf}$-like
fragments denote these variables. We can determine neutron and proton
numbers of these fragments, for the event $\lambda$, by integrating the particle
density on the left side or the right side of the dividing window. During contact, as a result
of nucleon flux across the window, neutron and proton numbers of these fragments
fluctuate from one event to another, and these numbers can be decomposed as 
$N_1^{\lambda}(t)=N_1(t)+\mathit{\delta N}_1^{\lambda }(t)$ and $Z_1^{\lambda
}(t)=Z_1(t)+\mathit{\delta Z}_1^{\lambda }(t)$, where $N_1(t)$ and  $Z_1(t)$
are the mean values taken over an ensemble of SMF events. For small amplitude
fluctuations, these mean values are determined by the mean-field description of
the TDHF theory. According to quantal diffusion approach, small amplitude
fluctuations of the neutron $\mathit{\delta N}_1^{\lambda }(t)$ and proton 
$\mathit{\delta Z}_1^{\lambda }(t)$ numbers evolve as a coupled linear
quantal Langevin 
equations~\cite{ayik2017,ayik2018,yilmaz2018,ayik2019,ayik2019b,sekizawa2020,ayik2020b}
\begin{align}\label{eq1}
\frac{d}{dt} \left(\begin{array}{c} {\delta Z_{1}^{\lambda} (t)} \\ {\delta N_{1}^{\lambda} (t)}
\end{array}\right)=&\left(\begin{array}{c} {\frac{\partial v_{p} }{\partial
Z_{1} } \left(Z_{1}^{\lambda } -Z_{1} \right)+\frac{\partial v_{p} }{\partial
N_{1} } \left(N_{1}^{\lambda } -N_{1} \right)} \\ {\frac{\partial v_{n}
}{\partial Z_{1} } \left(Z_{1}^{\lambda } -Z_{1} \right)+\frac{\partial v_{n}
}{\partial N_{1} } \left(N_{1}^{\lambda } -N_{1} \right)}
\end{array}\right)\nonumber\\ &+\left(\begin{array}{c} {\delta v_{p}^{\lambda }
(t)} \\ {\delta v_{n}^{\lambda } (t)} \end{array}\right).
\end{align}
\begin{figure*}[!th]
\includegraphics*[width=15cm]{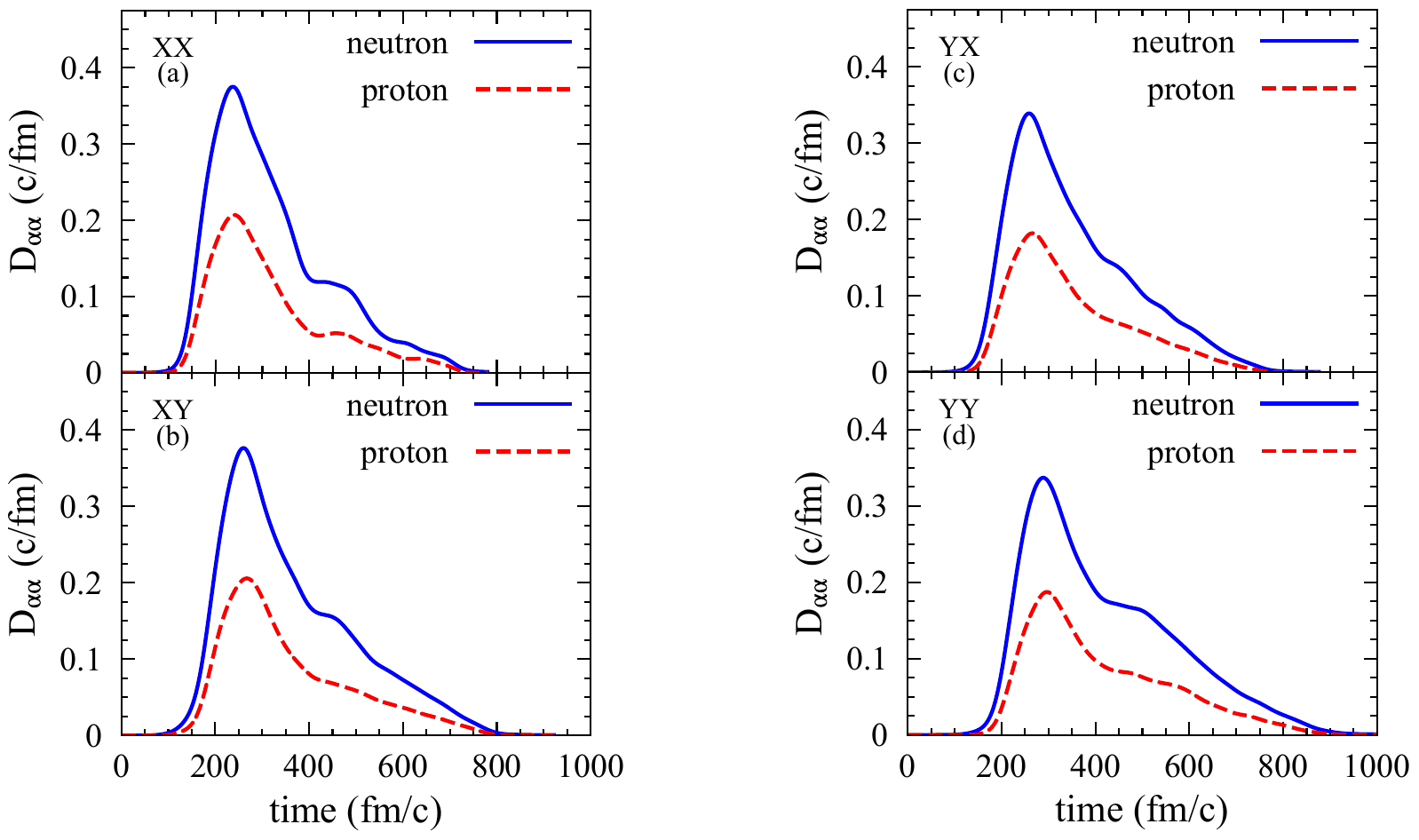}
\caption{Neutron and proton diffusion coefficient in the head-on collision of
${}^{250}\text{Cf}+{}^{232}\text{Th}$ system at $E_\text{c.m.}=950$~MeV  in tip-tip
(XX), tip-side (XY), side-tip (YX) and side-side (YY) collision geometries.
}
\label{fig2}
\end{figure*}
Quantities 
$v_{\alpha}^{\lambda}(t)=v_{\alpha}(t)+\delta v_{\alpha}^{\lambda }(t)$
are the drift coefficients of neutrons and protons
with the mean values and the fluctuating parts denoted by $v_{\alpha }(t)$
and $\mathit{\delta v}_{\alpha }^{\lambda }(t)$, respectively, with index $\alpha $
indicating neutron and proton labels. Drift coefficients $v_{\alpha }^{\lambda}(t)$
represent the rate of neutron and proton flux across the window for the
event $\lambda$. The linear limit of Langevin description presented here
provides a good approximation when the driving potential energy is nearly harmonic
around the equilibrium values of the mass and charge asymmetry. The mean values of the
drift coefficients are determined from the rate of change of neutron and proton
numbers in $\text{Cf}$-like fragments, which are shown in Fig.~\ref{fig1A} in
Appendix~\ref{appA}. The explicit quantal expressions of the stochastic parts of the drift
coefficients $\delta v^{\lambda}_{\alpha} (t)$ can be found in Ref.~\cite{ayik2017}.

\subsection{Quantal Diffusion Coefficients}

Stochastic part of the drift coefficients $\delta v^{\lambda}_{\text{p}} (t)$
and $\delta v^{\lambda}_{\text{n}}(t)$ provide the source for generating
fluctuations in mass and charge asymmetry degrees of freedom. According to the
SMF approach, stochastic parts of drift coefficients have Gaussian random
distributions with zero mean values $\delta \overline{v}_p^{\lambda }(t)=0$,
$\delta \overline{v}_n^{\lambda }(t)=0$, and the auto-correlation functions of
stochastic drift coefficient integrated over the history determine diffusion
coefficients $D_{\alpha\alpha} (t)$ for proton and neutron transfers,
\begin{align}\label{eq2}
\int_{0}^{t}dt'\overline{\delta v_{\alpha }^{\lambda } (t')\delta v_{\alpha
}^{\lambda}(t')} =D_{\alpha \alpha }
(t).                                                                           
\end{align}

In general diffusion coefficients involve a complete set of particle-hole
states. It is possible to eliminate the entire set of particle states by employing
closure relations in the diabatic limit. This results in an important simplification and as
a result, diffusion coefficients are determined entirely in terms of the
occupied single-particle wave functions of TDHF evolution. Explicit expressions
of the diffusion coefficients are provided in previous 
publications~\cite{ayik2017,ayik2018,yilmaz2018,ayik2019,ayik2019b,sekizawa2020,ayik2020b}
and for the
analysis of these coefficients please see Appendix~B in Ref.~\cite{ayik2017}. The fact
that diffusion coefficients are determined by the mean-field properties is
consistent with the fluctuation dissipation theorem of non-equilibrium
statistical mechanics and it greatly simplifies calculations of quantal
diffusion coefficients. Diffusion coefficients include quantal effects due
to shell structure, Pauli blocking, and the full effect of the collision geometry
without any adjustable parameters. We observe that there is a close analogy
between the quantal expression and the classical diffusion coefficient for a
random walk problem~\cite{gardiner1991,weiss1999,risken1996}.
The direct part is given as the sum of the nucleon
currents across the window from the target-like fragment to the projectile-like
fragment and from the projectile-like fragment to the target-like fragment,
which is integrated over the memory. This is analogous to the random walk
problem, in which the diffusion coefficient is given by the sum of the rate of
the forward and backward steps. The second part in the quantal diffusion
expression stands for the Pauli blocking effects in nucleon transfer mechanism,
which does not have a classical counterpart. As examples, Fig.~\ref{fig2} shows
the neutron and proton diffusion coefficients in head-on collisions of
${}^{250}\text{Cf}+{}^{232}\text{Th}$ system at $E_{\text{c.m.}}=950$~MeV for
different collision geometries. 
\begin{figure*}[!th]
\includegraphics*[width=15cm]{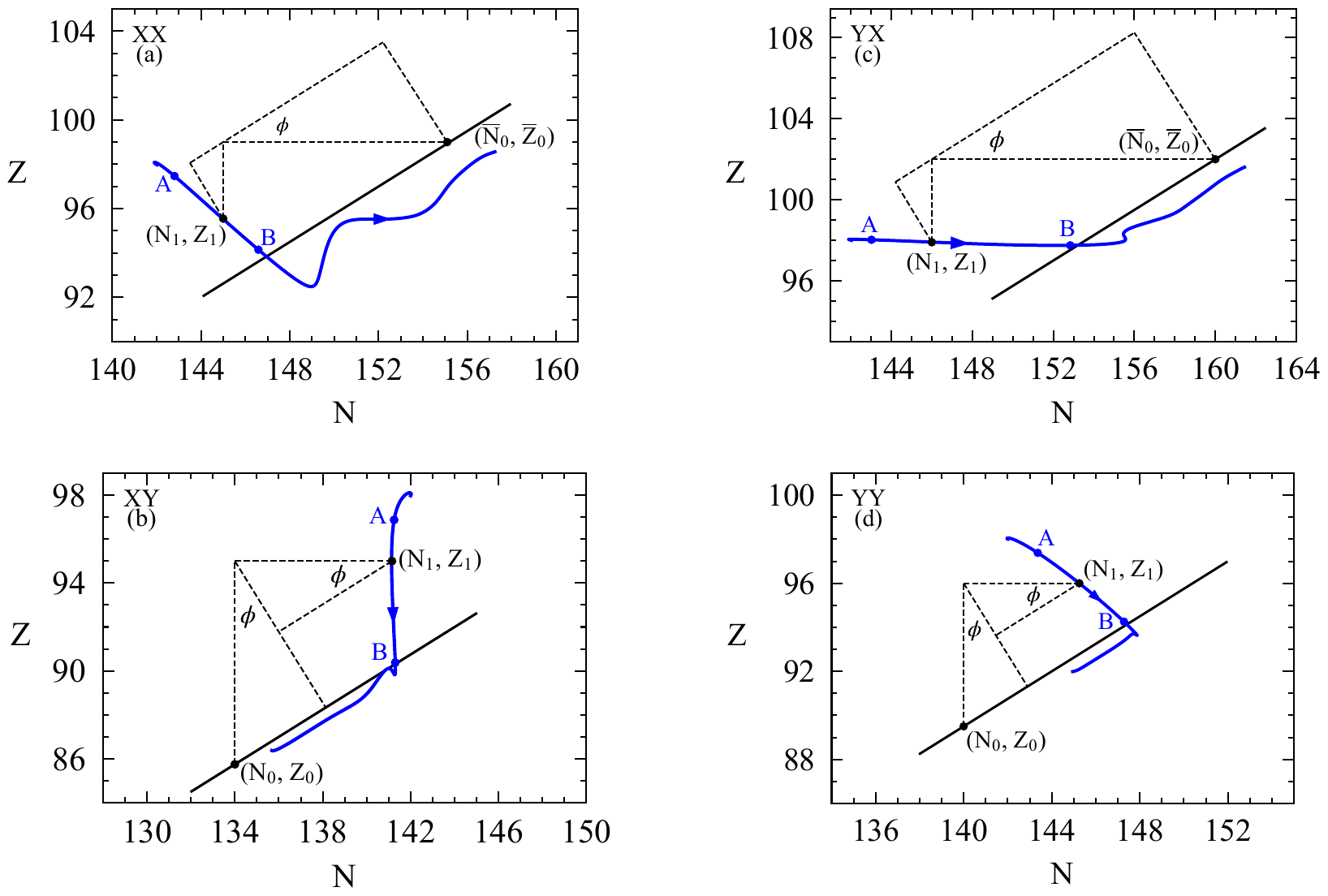}
\caption{Blue curves show drift path of Cf-like fragments in the head-on
collision of ${}^{240}\text{Cf}+{}^{246}\text{Th}$ system at $E_\text{c.m.}=950$~MeV in
tip-tip (XX), tip-side (XY), side-tip (YX) and side-side (YY) geometries.}
\label{fig3}
\end{figure*}

\subsection{Potential energy of di-nuclear system}
After colliding nuclei form a di-nuclear system, nucleon drift mechanism is
determined via the potential energy surface in the $(N,Z)${}-plane. Potential energy of
the di-nuclear system $U(N_1,Z_1)$ primarily consists of the surface energy, electrostatic
energy, symmetry energy and the centrifugal potential energy. TDHF theory includes
different energy contributions microscopically. Furthermore, TDHF
calculations show that the potential energy depends on the collision geometry. 
To compute the coupled Langevin equations, in addition to the diffusion
coefficients, we need to evaluate derivatives of the mean drift coefficients
with respect to neutron and proton numbers. The Einstein relations in the overdamped
limit~\cite{ayik2017,ayik2018,yilmaz2018,ayik2019,ayik2019b,sekizawa2020,ayik2020b}
provide a convenient approach to determine these derivatives. In the
over damped limit, drift coefficients are related to the potential energy
surface in the $(N,Z)${}-plane as,
\begin{subequations}
\begin{align}
 v_{n}(t)  &=-\frac{D_{NN}(t) }{T^{*} } \frac{\partial }{\partial N_{1} } U(N_1,Z_1)\;,\label{eq3a}\\
 v_{z}(t) &=-\frac{D_{ZZ}(t) }{T^{*} } \frac{\partial }{\partial Z_{1}} U(N_1,Z_1)\;,\label{eq3b}
\end{align}
\end{subequations}
where $T^{\ast }$ represents the effective temperature of the system. In heavy
di-nuclear systems, the centrifugal potential energy is not very important.
Therefore, we ignore the centrifugal potential energy and analyze potential
energy of di-nuclear systems formed in head-on collisions with  $\ell=0$ \ relative
angular momentum. In four geometries that we consider, di-nuclear system drifts
along isoscalar path toward a local equilibrium state. For collisions of
actinide nuclei, lighter local equilibrium state is located in vicinity of lead
valley with neutron and proton numbers around  $N_0=128$, $Z_0=82$ and heavier
local equilibrium state is located in vicinity of superheavy valley with neutron
and proton numbers around $\overline{N}_0=N_T-N_0=166$, 
$\overline{Z}_0=Z_T-Z_0=106$. Here $N_T =152+142$ total neutron number and
$Z_T=98+90$ total proton number of di-nuclear system, respectively. These nuclei
are located on the isoscalar path with the charge asymmetry $0.22$. As
illustrated in Fig.~\ref{fig1}, di-nuclear system formed in collision of
$^{250}\text{Cf}+^{232}\text{Th}$ drifts nearly along the isoscalar path, which
is parallel to the equilibrium valley of stable nuclei. To extract information
about potential energy in perpendicular direction to stability valley, we need
to choose reaction of a suitable neighboring system. For this purpose, head-on
collision of $^{240}\text{Cf}+^{246}\text{Th}$ system provides a suitable
system. Fig.~\ref{fig3} shows drift path of Cf-like fragments in head-on
collision of $^{240}\text{Cf}+^{246}\text{Th}$ reaction at tip-tip (XX),
tip-side (XY), side-tip (YX) and side-side (YY) geometries at 
$E_{\text{c.m.}}=950$~MeV. In Appendix~\ref{appA}, Fig.~\ref{fig2A}  shows neutron and
proton numbers of Cf-fragments as function of time in different geometries. In
this system, the initial charge asymmetry of $^{240}{\text{Cf}}$ is
$(N-Z)/(N+Z)=0.18$ and of $^{246}{\text{Th}}$ is $(N-Z)/(N+Z)=0.27$. Initially,
system rapidly drifts towards equilibrium valley, until reaches equilibrium
charge asymmetry value 0.22. The perpendicular component of this drift line is
referred to as the isovector path. Then, system continues to drift along
isoscalar path toward symmetry or asymmetry with the same charge asymmetry and
the same slope angle as the isoscalar path in $^{250}\text{Cf}+^{232}\text{Th}$
system. System separates before reaching local equilibrium. Combining the drift
information of these two very similar systems, we can provide an approximate
description of potential energy surface of di-nuclear system relative to the
equilibrium value in terms of two parabolic forms~\cite{merchant1982}, 
\begin{align}\label{eq4}
U(N_{1} ,Z_{1} )=\frac{1}{2} aR_{S}^{2} (N_{1} ,Z_{1} )+\frac{1}{2} bR_{V}^{2} (N_{1} ,Z_{1} )\;.
\end{align} 
Here, $R_S(N_1,Z_1)$ and  $R_V(N_1,Z_1)$ represent perpendicular distances of a
fragment with neutron and proton numbers $(N_1,Z_1)$ from the isoscalar path
and from the local equilibrium state along the isoscalar path, respectively.
Because of the sharp increase of asymmetry energy, we expect the isovector
curvature parameter $a$ to be much larger than the isoscalar curvature parameter
$b$. When drift occurs toward symmetry like in Fig.~\ref{fig3}(d), we can
express the isoscalar distance from the local equilibrium state $(N_0,Z_0)$ as,
$R_V=(N_1-N_0)\text{cos}\phi +(Z_1-Z_0)\text{sin}\phi$, and the isovector distance from the
isoscalar path as $R_S=(Z_1-Z_0)\text{cos}\phi -(N_1-N_0)\text{sin}\phi $. When
drift occurs toward asymmetry like in Fig.~\ref{fig3}(c), we can express
isoscalar distance from local the equilibrium state  $(\overline{N}_0,
\overline{Z}_0)$ as $\overline{R}_V=(\overline{N}_0-N_1)\text{cos}\phi
+(\overline{Z}_0-Z_1)\text{sin}\phi $, and the isovector distance as
$\overline{R}_S=(\overline{N}_0-N_1)\text{sin}\phi
-(\overline{Z}_0-Z_1)\text{cos}\phi$. The angle  $\phi$ is the angle between the
isoscalar path and  $N-$ axis, which is about $\phi =32^{\circ}$. It is
possible to derive similar expressions for other geometries. All have the same
isoscalar path which makes the same angle  $\phi =32^{\circ}$ with N-axis.
Because of analytical relations of the potential energy in the Einstein
relations, we can immediately calculate derivatives of drift coefficients to
find,
\begin{align}
\frac{\partial \nu _{n}}{\partial N_{1} } &=-D_{NN} \left(\alpha \sin ^{2} \phi +\beta \cos ^{2} \phi \right)\label{eq5}\;,\\
\frac{\partial \nu _{z} }{\partial Z_{1} } &=-D_{ZZ} \left(\alpha \cos ^{2} \phi +\beta \sin ^{2} \phi \right)\label{eq6}\;,\\
\frac{\partial \nu _{n} }{\partial Z_{1} } &=-D_{NN} \left(\beta -\alpha \right)\sin \phi \cos \phi\label{eq7}\;,\\
\frac{\partial \nu _{z} }{\partial N_{1} } &=-D_{ZZ} \left(\beta -\alpha \right)\sin \phi \cos \phi\label{eq8}\;.
\end{align} 
These expressions are valid for different collision geometries with different
values of reduced isoscalar, $\beta =b/T^{\ast}$, and isovector, $\alpha
=a/T^{\ast}$, curvature parameters. By inverting Eq.~(\ref{eq3a}) and Eq.~(\ref{eq3b}),
it is possible to express the reduced curvature parameters in
terms of drift and diffusion coefficients. Due to microscopic shell structure,
transport coefficients depend on time, as a result, the reduced curvature
parameters are time dependent as well. In simple parabolic parametrization of
potential energy surface, we ignore the time dependence and use constant
curvature parameters. Constant curvature parameters are determined by averaging
over suitable time intervals while colliding nuclei have sufficiently large
overlap. When drift occurs toward symmetry, the averaged value of the isoscalar
reduced curvature parameter over a time interval $t_1$ and $t_2$ is
determined as,
\begin{align}\label{eq9}
\beta(12) = -\frac {1} {R_{V} (12)} \int _{t_1}^{t_2}\left(\frac{v_{n} (t)\cos
\phi }{D_{NN} (t)} +\frac{v_{p} (t)\sin \phi }{D_{ZZ} (t)} \right)  dt\;,
\end{align} 
where the integrated isoscalar distance is given by
\begin{align}\label{eq10}
R_V(12)=\int _{t_1}^{t_2}\left\{(N_1(t)-N_0)\text{cos}\phi +(Z_1(t)-Z_0)\text{sin}\phi
\right\} dt\;.
\end{align} 
We can use these expressions in calculating averaged values of isoscalar
reduced curvature parameters in different geometries. In tip-tip collision of
${}^{250}\text{Cf}+{}^{232}\text{Th}$ system, we take the averaging interval, as
shown in Fig.~\ref{fig1A}(a) in Appendix~\ref{appA}, as  $t_1 \rightarrow t_A=150$~fm/c,
and $t_2 \rightarrow t_B=550$~fm/c, we find the reduced isoscalar curvature
parameter to be $\beta (\text{tt})=0.005$. In tip-side geometry using the
interval as  $t_1 \rightarrow t_A=200$~fm/c and $t_2\rightarrow t_B=500$~fm/c, as
shown in Fig.~\ref{fig1A}(b) of Appendix~\ref{appA}, we find the reduced isoscalar
curvature parameter to be $\beta (\text{ts})=0.009$. When drift is toward
asymmetry, we can determine the averaged value of the isoscalar reduced
curvature parameter over a time interval $t_1$ and  $t_2$ using the
negative of expression (\ref{eq9}) and by taking the integrated isoscalar
distance as  $R_V(12) \rightarrow \overline{R}_V(12)$,
\begin{align}\label{eq11}
\overline{R}_V(12)=\int _{t_1}^{t_2}\left\{(\overline{N}_0-N_1
(t))\text{cos}\phi +(\overline{Z}_0-Z_1(t))\text{sin}\phi \right\} dt\;.
\end{align} 

In side-tip geometry, using the interval $t_A=200$~fm/c and $t_B=500$~fm/c, as
shown in Fig.~\ref{fig1A}(c) in Appendix~\ref{appA}, we find the reduced isoscalar
curvature parameter to be $\beta ({\text{st}})=0.009$. For the side-side geometry, we
estimate the isoscalar curvature parameters in the interval $t_A \rightarrow
t_B$ with $t_A=200$~fm/c and $t_B=300$~fm/c, as shown in Fig.~\ref{fig1A}(d) in
Appendix~\ref{appA}, to be $\beta
({\text{ss}})=0.004$. In the interval $t_B \rightarrow t_C$ with $t_B=300$~fm/c
and $t_C=800$~fm/c, we estimate the isoscalar curvature parameter to have the
same magnitude, $\beta ({\text{ss}})=0.004$.

When drift occurs toward symmetry, we estimate the isovector reduced curvature
parameters in different collision geometries from the drift paths of 
${}^{240}\text{Cf}+{}^{246}\text{Th}$ by averaging over time interval  $t_1$ and  $t_2$ as,
\begin{align}\label{eq12}
\alpha(12) = \frac {1} {R_{S} (12)} \int _{t_1}^{t_2}\left(\frac{v_{n} (t)\sin
\phi }{D_{NN} (t)} -\frac{v_{p} (t)\cos \phi }{D_{ZZ} (t)} \right)  dt\;,
\end{align}
where the integrated isovector distance is given by
\begin{align}\label{eq13}
R_S(12)=\int _{t_1}^{t_2}\left\{(Z_1(t)-Z_0)\text{cos}\phi -(N_1(t)-N_0)\text{sin}\phi
\right\}dt\;.
\end{align}
\begin{figure*}[!th]
\includegraphics*[width=15cm]{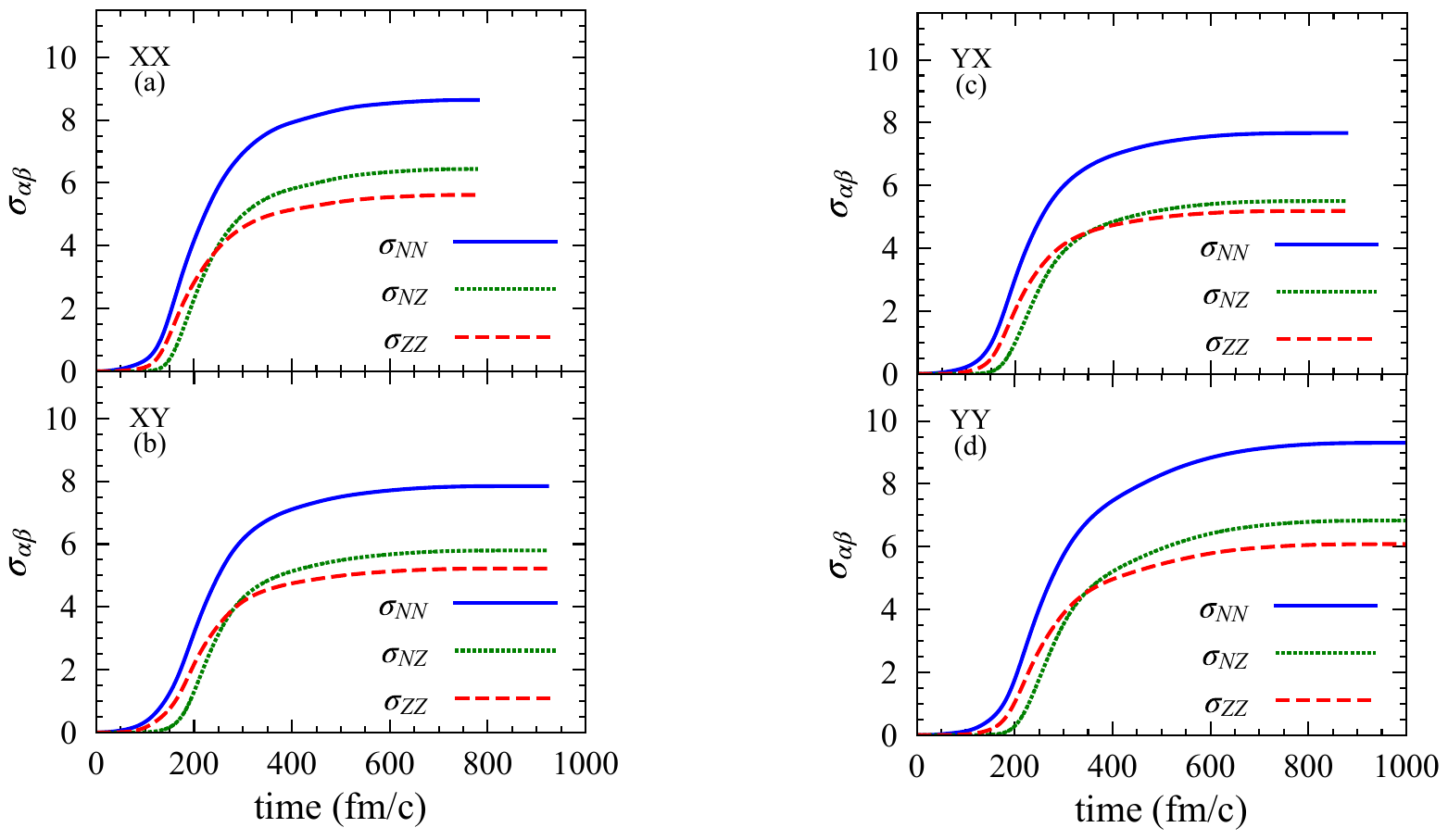}
\caption{Neutron, proton and mixed variances as a function of time in the
${}^{250} \text{Cf}+{}^{232} \text{Th}$ system at $E_\text{c.m.}=950$~MeV  in tip-tip
(XX), tip-side (XY), side-tip (YX) and side-side (YY) geometries. }
\label{fig4}
\end{figure*}

In side-side geometry, we estimate the isovector curvature parameter by
calculating the average value in the interval  $t_A \rightarrow t_B$ with 
$t_A=220$~fm/c and $t_B=310$~fm/c, as shown by Fig.~\ref{fig2A}(d) in Appendix~\ref{appA}.
We find the reduced isoscalar curvature parameter to be
$\alpha({\text{ss}})=0.11$. In tip-side geometry, we estimate the isovector
curvature parameter by calculating the average value in the interval $t_A
\rightarrow t_B$ with $t_A=170$~fm/c and $t_B=280$~fm/c, as shown by 
Fig.~\ref{fig2A}(b) in Appendix~\ref{appA}. 
We find the reduced isoscalar curvature parameter
to be $\alpha ({\text{ts}})=0.15$. When drift occurs toward asymmetry, we
estimate the isovector reduced curvature parameters in different collision geometries
from the drift paths of ${}^{240}\text{Cf}+{}^{246}\text{Th}$ using 
Eq.~(\ref{eq12}) in which the integrated isovector distance is replaced by
$R_S(12) \rightarrow \overline{R}_S(12)$,
\begin{align}\label{eq14}
\overline{R}_S(12)=\int _{t_1}^{t_2}\left\{(\overline{Z}_0-Z_1
(t))\text{cos}\phi - (\overline{N}_0-N_1(t))\text{sin}\phi\right\} dt\;.
\end{align} 

In side-tip geometry, we estimate the isovector curvature parameter by
calculating the average value in the interval  $t_A \rightarrow t_B$ with 
$t_A=150$~fm/c and $t_B=260$~fm/c, as shown by Fig.~\ref{fig2A}(c) in Appendix~\ref{appA}.
We find the reduced isoscalar curvature parameter to be
$\alpha({\text{st}})=0.12$. In tip-tip geometry, we estimate the isovector
curvature parameter by calculating the average value in the interval $t_A
\rightarrow t_B$ with  $t_A=150$~fm/c and $t_B=220$~fm/c, as shown by Fig.
\ref{fig2A}(a) in Appendix~\ref{appA}. We find the reduced isoscalar curvature parameter
to be $\alpha ({\text{tt}})=0.16$. 

\section{Primary and secondary cross-sections of reaction products}
\label{sec4}
\subsection{Probability distributions of primary fragments}

The joint probability distribution function $P_\ell(N,Z)$ for producing binary
fragments with $N$ neutrons and $Z$ protons is determined by generating a
large number of solutions of the Langevin Eq.~(\ref{eq1}). It is well known that the
Langevin equation is equivalent to the Fokker-Planck equation for the
distribution function of the macroscopic variables~\cite{gardiner1991}.
In the special case,
when drift coefficients are linear functions of macroscopic variables, as we
have in Eq.~(\ref{eq1}), the proton and neutron distribution functions for the
initial orbital angular momentum $\ell$ is given as a correlated Gaussian function
described by the mean values, neutron, proton and mixed dispersions as, 
                          
\begin{align}\label{eq15}
P_{\ell} (N,Z)=\frac{1}{2\pi \sigma _{NN}(\ell)\sigma _{ZZ}(\ell)\sqrt{1-\rho
_{\ell}^{2}}}\exp\left(-C_{\ell}\right)\;.
\end{align}
Here, the exponent $C_l$ for each initial angular momentum is given by
\begin{align}\label{eq16}
C_{\ell} =\frac{1}{2\left(1-\rho_{\ell}^{2} \right)}
&\left[\left(\frac{Z-Z_{\ell} }{\sigma _{ZZ} (\ell)} \right)^{2} -2\rho_\ell
\left(\frac{Z-Z_{\ell} }{\sigma _{ZZ} (\ell)} \right)\left(\frac{N-N_{\ell}
}{\sigma _{NN} (\ell)} \right)\right.\nonumber\\ &\left.+\left(\frac{N-N_{\ell}
}{\sigma _{NN} (\ell)} \right)^{2} \right]\;,
\end{align}
with the correlation coefficient defined as $\rho _\ell=\sigma
_{\text{NZ}}^2(\ell)/(\sigma _\text{ZZ}(\ell)\sigma _{\text{NN}}(\ell))$.
Quantities $N_\ell=\overline{N}_\ell^{\lambda }$, 
$Z_\ell=\overline{Z}_\ell^{\lambda }$ denote the mean neutron and proton numbers
of the target-like or project-like fragments. These mean values are determined
from the TDHF calculations.
It is possible to
deduce coupled differential equations for variances $\sigma
_{\text{NN}}^2(\ell)=\overline{\mathit{\delta N}^{\lambda}\mathit{\delta
N}^{\lambda }}$,  $\sigma _{\text{ZZ}}^2(\ell)=\overline{\mathit{\delta Z}^{\lambda
}\mathit{\delta Z}^{\lambda }}$, and co-variances  $\sigma
_{\text{NZ}}^2(\ell)=\overline{\mathit{\delta N}^{\lambda }\mathit{\delta
Z}^{\lambda }}$ by multiplying Langevin Eq.~(\ref{eq1}) with $\mathit{\delta
N}^{\lambda }$, ${\delta Z}^{\lambda }$ and carrying out the average over the
ensemble generated from the solution of the Langevin equation. These coupled
equations were presented in 
Refs.~\cite{ayik2017,ayik2018,yilmaz2018,ayik2019,ayik2019b,sekizawa2020,ayik2020b}.
For completeness, we provide these
differential equations here~\cite{schroder1981},
\begin{align}\label{eq17}
\frac{\partial}{\partial t } {\sigma}^2_{NN} = 2 \frac{\partial
\nu_{n}}{\partial N_1} \sigma^2_{NN} + 2 \frac{\partial \nu_{n}}{\partial
Z_1}\sigma^2_{NZ} + 2 D_{NN}\;,
\end{align}
\begin{align}\label{eq18}
\frac{\partial}{\partial t } {\sigma}^2_{ZZ} = 2 \frac{\partial
\nu_{p}}{\partial Z_1} \sigma^2_{ZZ} + 2 \frac{\partial \nu_{p}}{\partial
N_1}\sigma^2_{NZ} + 2 D_{ZZ}\;,
\end{align}
and
\begin{align}\label{eq19}
\frac{\partial}{\partial t } {\sigma}^2_{NZ} =  \frac{\partial \nu_{p}}{\partial
N_1} \sigma^2_{NN} +  \frac{\partial \nu_{n}}{\partial Z_1}\sigma^2_{ZZ} +
\sigma^2_{NZ}\left(\frac{\partial \nu_p}{\partial Z_1} +\frac{\partial
\nu_n}{\partial N_1}\right).
\end{align}
Here, $D_{NN}$ and $D_{ZZ}$ indicate the diffusion coefficients for proton and
neutron transfer. Variances and co-variances are determined from the solutions
of these coupled differential equations with initial conditions $\sigma
_{NN}^2(t=0)=0$,  $\sigma_\text{NN}^2(t=0)=0$ and  $\sigma_\text{NN}^2(t=0)=0$,
for each orbital angular momentum. As an example, Fig.~\ref{fig4} shows neutron, proton,
and mixed variances as a function of time in head-on collision of
${}^{250}{\text{Cf}}+{}^{232}{\text{Th}}$ at  $E_{\text{c.m.}}=950$~MeV for the tip-tip collision
geometry.

\begin{figure*}[!ht]
\includegraphics*[width=17.5cm]{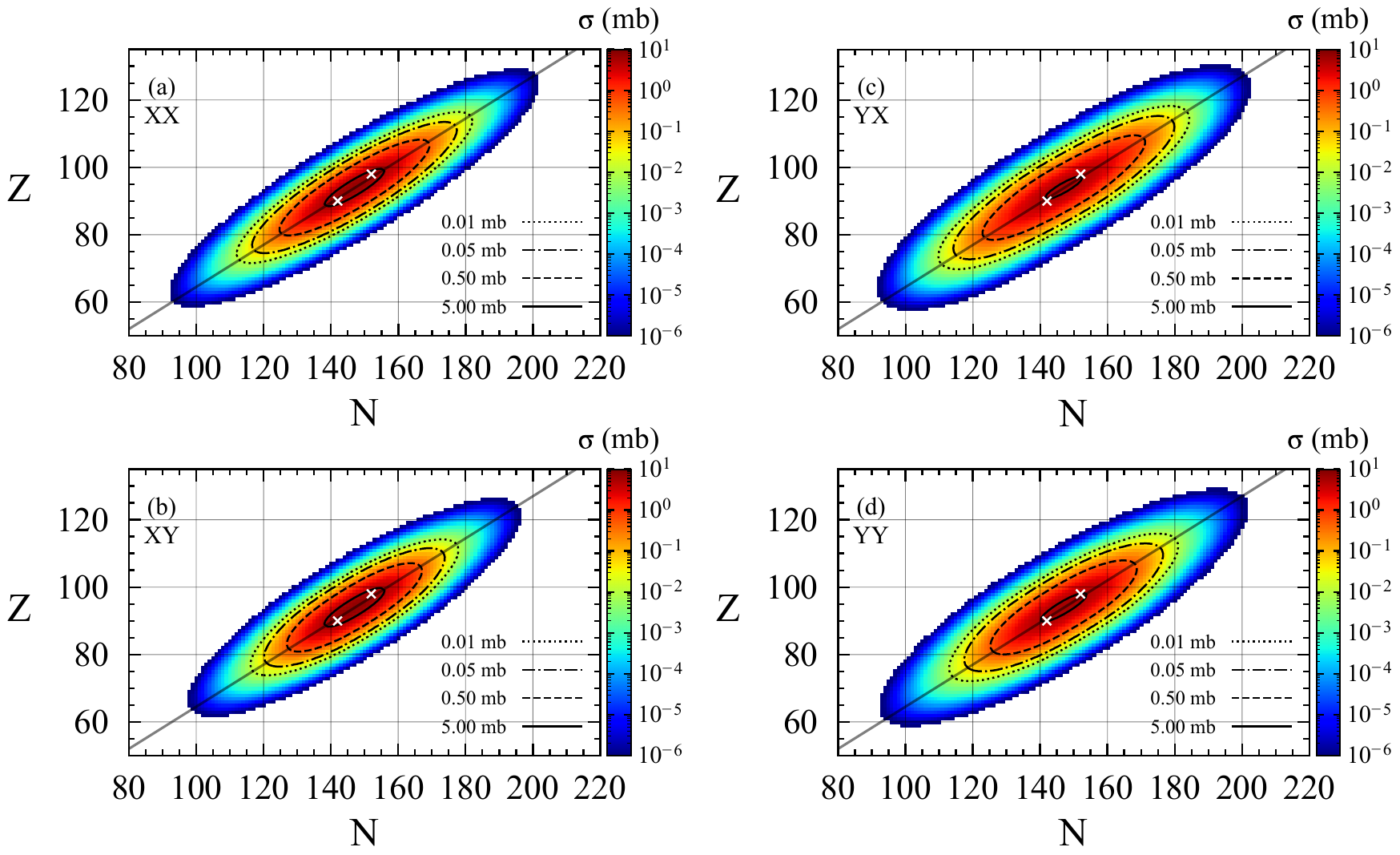}
\caption{Primary production cross sections in the NZ plane for 
${}^{250}\text{Cf}+{}^{232}\text{Th}$ system at $E_\text{c.m.}=950$~MeV  in tip-tip
(XX), tip-side (XY), side-tip (YX) and side-side (YY) geometries.}
\label{fig5}
\end{figure*}
\subsection{Cross-sections of primary reaction products}

We calculate the cross sections for production of primary isotopes using the standard expression,
\begin{align}\label{eq20}
\sigma ^{pri}(N,Z)=\frac{\pi \hbar ^2}{2{\mu}E_\text{c.m.}}\sum
\limits_{\ell_{min}}^{\ell_{max}} (2\ell +1 )P^{pri}_{\ell}(N,Z)\;,
\end{align}
where
\begin{align}\label{eq21}
P_{\ell}^{pri}(N,Z)=\frac{1}{2}[P_{\ell}^{pro}(N,Z)+P_{\ell}^{tar}(N,Z)]\;.
\end{align}
In this expression,  $P_{\ell}^{pro}(N,Z)$ and $P_{\ell}^{tar}(N,Z)$ denote the
normalized probability of producing projectile-like and target-like fragments.
These probabilities are given by Eq.~(\ref{eq15}) using mean values of
projectile-like and target-like fragments, respectively. The factor of $1/2$ is
introduced to make the total primary fragment distribution normalized to unity.
In summation over $\ell$, the range of initial orbital angular momenta depend on the
detector geometry in the laboratory frame. There is no nucleon transfer data
available for the $^{250}\text{Cf}+^{232}\text{Th}$ system. In calculations, we
carry out summation over the range from $\ell_{min}=0\hbar$ to
$\ell_{max}=480\hbar$. The upper limit corresponds to quasi elastic scattering
events with a few nucleons transfer channels. We calculate total double
cross-sections for four different tip-tip, tip-side, side-tip and side-side collision
geometries. Fig.~\ref{fig5} shows the double cross-sections in (N-Z) plane for the
production of primary fragments in tip-tip (a), tip-side (b), side-tip (c), and
side-side (d) geometries. The points shown by crosses indicate colliding nuclei
$^{250}\text{Cf}$ and $^{232}\text{Th}$. Equal values of primary cross-sections
form elliptic curves. Large values of mixed dispersions indicate strong
correlations in neutron-proton transfers. The strong correlations are induced
mainly by the symmetry energy. As a result, the major axes of equal
cross-section elliptic curves are aligned along with the valley of stability. Gross
properties of primary cross-sections are similar in different collision geometries.
Due to the drift towards asymmetry direction, magnitude of the cross-sections along the
isoscalar direction extends further towards the super heavy-island in side-tip
collision geometry as compared to the other collision geometries. As an example,magnitude of primary
cross-section for production of element (Z=118, N=185) is about 0.01 mb, which
is larger than those in other collision geometries. 

\subsection{Cross-sections of secondary reaction products}
\setcounter{topnumber}{1}
\begin{figure*}[!hbt]
\includegraphics*[width=17.5cm]{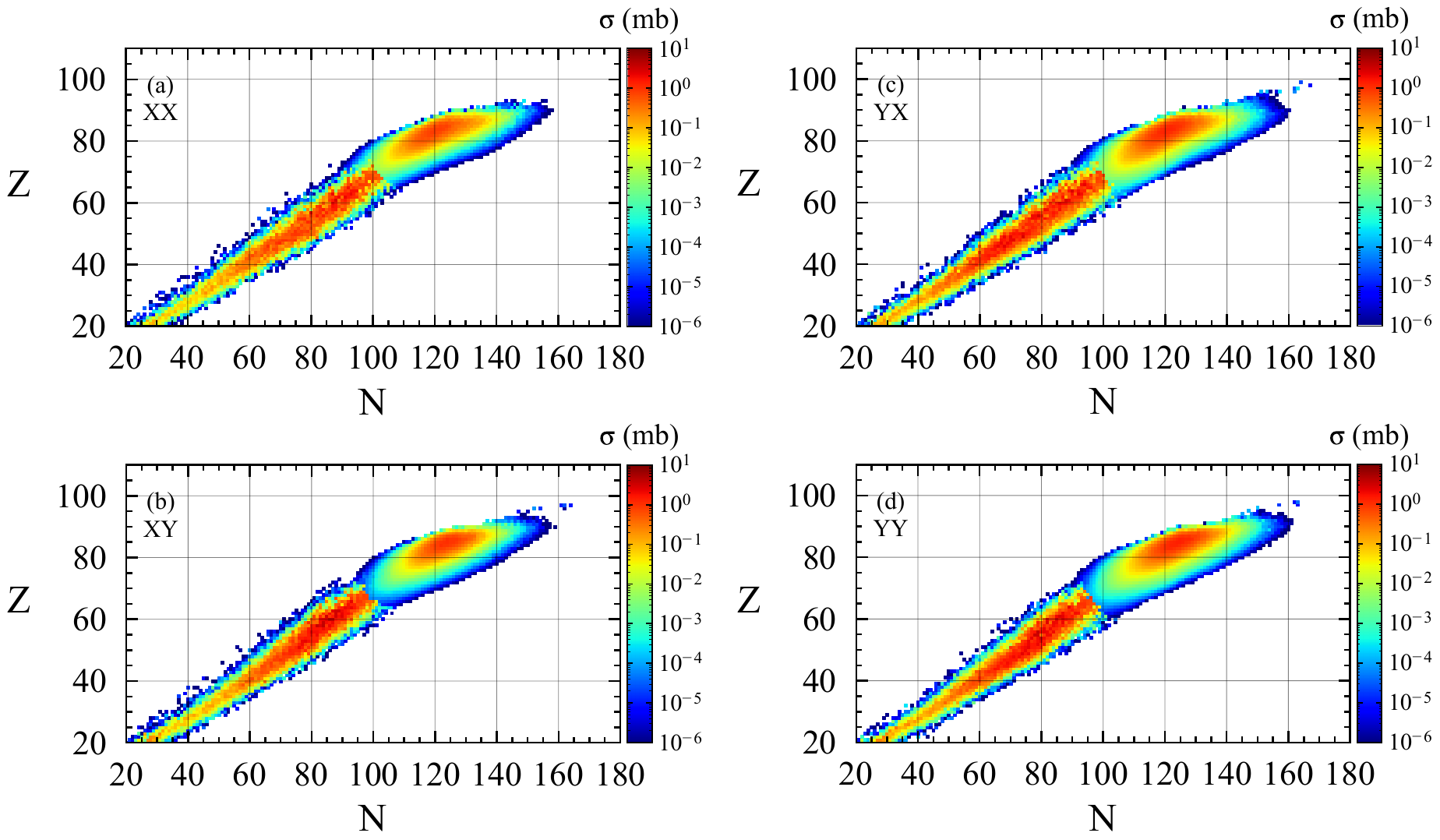}
\caption{Secondary production cross sections in the NZ plane for 
${}^{250}\text{Cf}+{}^{232}\text{Th}$ system at $E_\text{c.m.}=950$~MeV  in tip-tip
(XX), tip-side (XY), side-tip (YX) and side-side (YY) geometries.}
\label{fig6}
\end{figure*}
Primary fragments are excited and cool down by light particles emission,
mostly neutrons, protons, and alpha particles, or they may decay via binary fission. We
analyze the de-excitation mechanisms of the primary fragments using the statistical code
GEMINI++~\cite{charity2008}. We estimate the total excitation energy of the primary fragments
according to $E_{\ell}^{\ast}(Z,N)=E_\text{c.m.}-\text{TKE}_{\ell}-Q_{\text{gg}}(Z,N)$.
In this expression $\text{TKE}_{\ell}$ is the mean value of total asymptotic
kinetic energy in collision with initial orbital angular momentum $\ell$, and 
$Q_{\text{gg}}(Z,N)$ denotes ground state $Q$-value of the primary
fragments relative to the initial value. For collisions with an initial orbital angular
momentum, in the exit channel total spin and total excitation energy should have
distributions around their mean values.
\begin{figure*}[!th]
\includegraphics*[width=17.5cm]{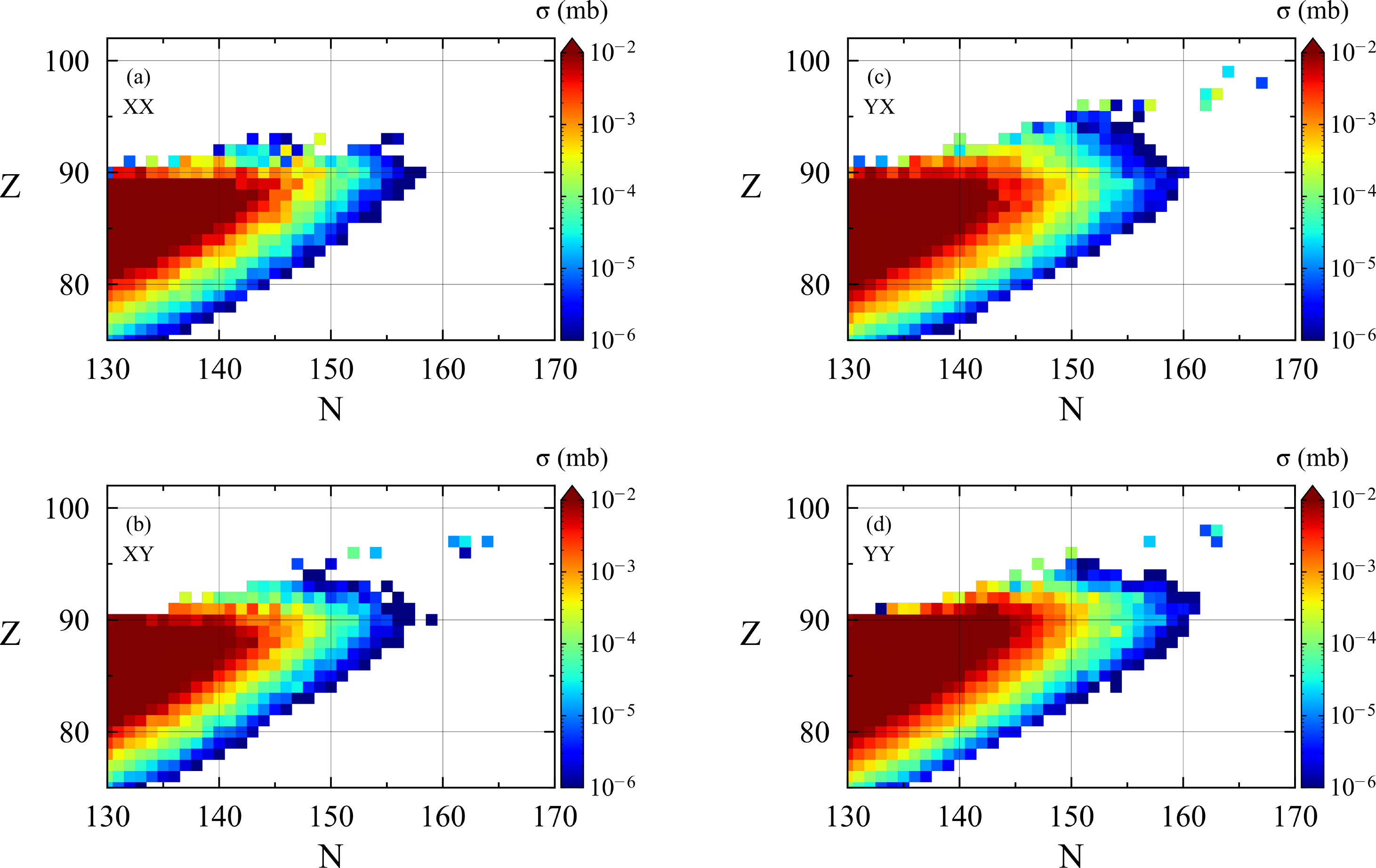}
\caption{Enlarged view of secondary production cross sections in the NZ plane
for  ${}^{250}\text{Cf}+{}^{232}\text{Th}$ system at $E_\text{c.m.}=950$~MeV  in
tip-tip (XX), tip-side (XY), side-tip (YX) and side-side (YY) geometries.
}
\label{fig7}
\end{figure*} 
In the present analysis, we ignore the fluctuations in excitation energy and
spin of primary fragments. We share the mean value of the total excitation energy and
the total angular momentum transfer in proportion to mass ratio of the primary
fragments. The excited parent nucleus decays by a series of particle emissions
and by secondary fission until the decay of parent nucleus is energetically
forbidden. Starting from an exited parent nucleus with neutron and proton
numbers $(Z,N)$, excitation energy $E^{\ast}(Z,N)$, and spin $J$, statistical
code GEMINI++ determines the probability $W(N,Z \rightarrow N',Z')$ of reaching
final nucleus $(Z',N')$. Probability distribution of secondary
fragments are specified as,
\begin{align}\label{eq22}
P_{\ell}^{sec}(N',Z')=\sum_{N\ge N'}\sum_{Z\ge Z'}P_{\ell}^{pri}(N,Z)W(N,Z \rightarrow N',Z')\;.
\end{align}
Here, summation $(Z,N)$ covers the pairs of projectile-like fragments and target-like
fragments of the di-nuclear system according to their probability distributions. 
\begin{align}\label{eq23}
\sigma_{\ell}^{sec}(N',Z')=\frac{\pi \hbar ^2}{2\mu
E_\text{c.m.}}\sum_{\ell_{min}}^{\ell_{max}}(2\ell+1)P_{\ell}^{sec}(N',Z')\;.
\end{align}
Figure~\ref{fig6} shows the double cross-sections of secondary fragments in the (N-Z)
plane for different collision geometries. Again, the gross properties of the secondary
cross-sections are very similar for different collision geometries. Below $N=100$, and
$Z=60$, decay products consist of secondary fission of excited heavy fragments.
Above this region the cross-sections are populated by light particle emission including neutron,
proton, and alpha particles. Calculations predict production of broad range of
neutron rich isotopes for nuclei with proton numbers in the range of $Z=70-90$,
with cross-sections on the order of several hundred  micro barns. Figure~\ref{fig7}
shows an enlarged view of the secondary cross-sections for heavy neutron rich nuclei for
the range of proton numbers $Z=80-90$, and range of neutron numbers
$N=130-160$. Furthermore, our calculations indicate a number of neutron rich heavy
isotopes with sizable cross-sections including $^{246}\text{Cm}$ with a
cross-section of 159 nanobarn for the side-side collision, $^{248}\text{Cm}$ with a
cross-section of 80 nanobarn for the tip-side collision, and $^{253}\text{Cm}$ with a
cross-section of 252 nanobarn for the side-tip collision.

\section{Conclusions}
\label{sec5}
We have presented an investigation of multinucleon transfer mechanism for the collisions of
$^{250}\text{Cf}+^{232}\text{Th}$ system at $E_\text{c.m.}=950$~MeV employing the quantal
transport description based on the SMF approach. The standard mean-field
description of TDHF determines average evolution of most probable path of
heavy-ion collision dynamics at low energies. The SMF provides an extension to the
standard TDHF description by including mean-field fluctuations in a manner
consistent with fluctuation-dissipation theorem of non-equilibrium statistical
mechanics. When a di-nuclear complex is maintained in collisions, we can extract
Langevin equations for macroscopic variables, such as the mass and charge asymmetry
of colliding ions. In this work, we take neutron and proton numbers of one of
the collision partners in the di-nuclear complex as the relevant macroscopic variables.
Using the equivalence of Langevin description and Fokker-Planck descriptions,
it is possible to provide nearly an analytical description in terms of correlated
Gaussian shape probability distribution of the primary fragments produced in
collisions. The correlated Gaussian distribution for each orbital angular momentum is
determined by the asymptotic values of the mean neutron and proton numbers of the primary
fragments, and the neutron, proton, and mixed dispersions. We determine these
dispersions employing the quantal transport approach.
Diffusion coefficients, which provide the source for
developing fluctuations, are evaluated in terms of the occupied single-particle wave
functions of the TDHF theory. Transport coefficients include quantal effects due to
shell structure and Pauli blocking, and do not involve any adjustable parameters
other than the standard parameters of the effective Skyrme force used in the TDHF
calculations. Highly excited primary fragments decay by particle emission and
secondary fission. Employing the statistical code GEMINI++, we can follow the de-excitation
process of primary fragments, and calculate the production cross-sections for the
secondary fragments. Since there is no data available, we are not able to test
our prediction for multinucleon transfer mechanism for the
$^{250}\text{Cf}+^{232}\text{Th}$ reaction.

\begin{acknowledgments}
S.A. gratefully acknowledges Middle East Technical University for warm
hospitality extended to him during his visits. S.A. also gratefully acknowledges
F. Ayik for continuous support and encouragement. This work is supported in part
by US DOE Grants Nos. DE-SC0015513 and DE-SC0013847.
This work is supported in part by TUBITAK Grant No. 122F150. The numerical
calculations reported in this paper were partially performed at TUBITAK ULAKBIM,
High Performance and Grid Computing Center (TRUBA resources).
\end{acknowledgments}

\appendix
\section{}
\label{appA}
We can estimate the averaged values of reduced isoscalar and isovector
curvature parameters with the help of Einstein relations, Eq.~(\ref{eq3a}) and 
Eq.~(\ref{eq3b}), in the overdamped limit. We evaluate the average values of the reduced
curvature parameters for different geometries by carrying out the time integrals in
Eq.~(\ref{eq9}) and Eq.~(\ref{eq12}) over suitable time intervals. These time
intervals are indicated in following figures for 
$^{250}\text{Cf}+^{232}\text{Th}$ and $^{240}\text{Cf}+^{246}\text{Th}$
reactions.

\begin{figure*}[!th]
\includegraphics*[width=14cm]{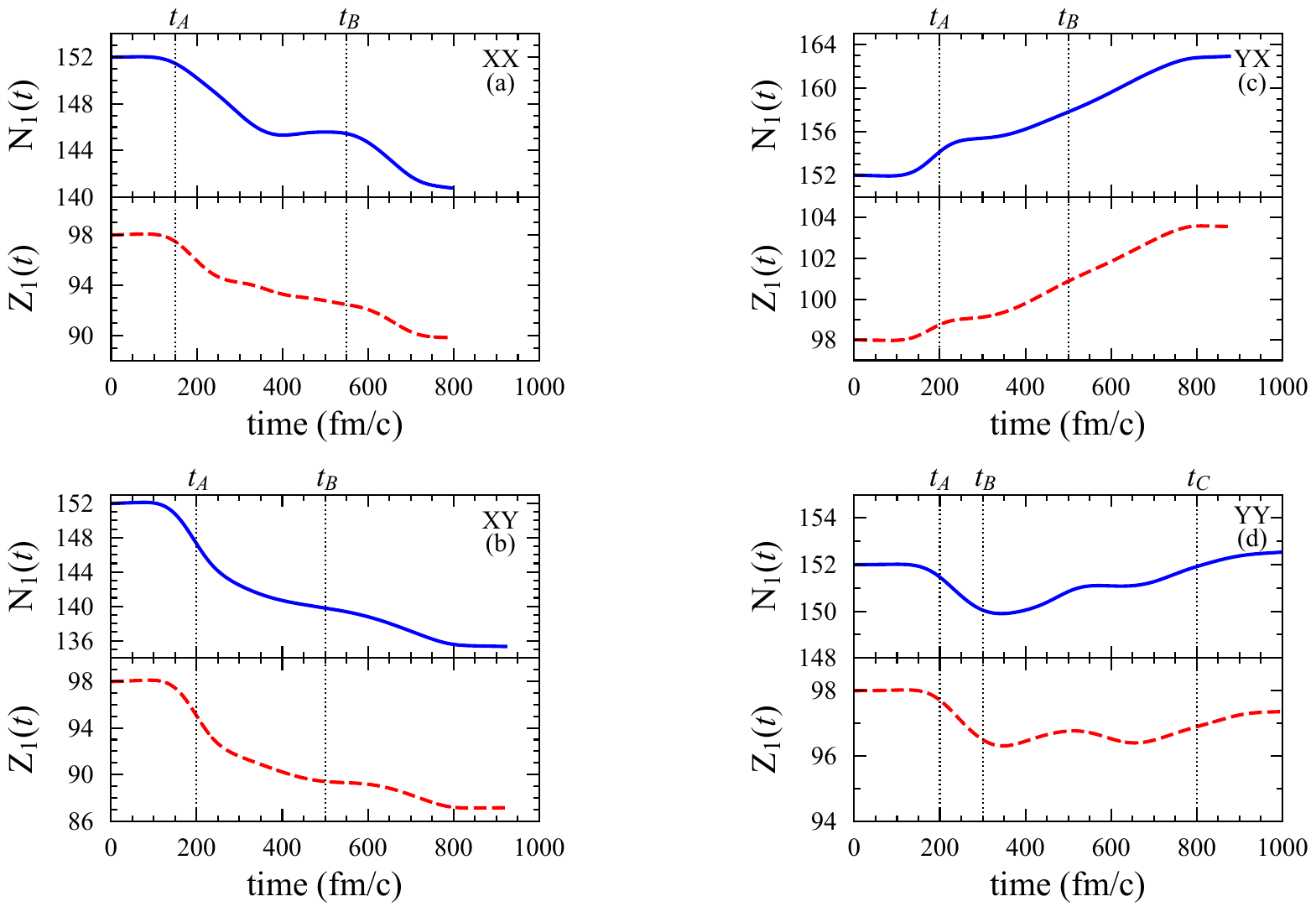}
\caption{Mean-values of neutron and proton numbers of Cf-like fragments in the
head-on collision of ${}^{250}\text{Cf}+{}^{232}\text{Th}$ system at
$E_\text{c.m.}=950$~MeV  in tip-tip (XX), tip-side (XY), side-tip (YX) and side-side
(YY) geometries.}
\label{fig1A}
\end{figure*}
\begin{figure*}[!th]
\includegraphics*[width=14cm]{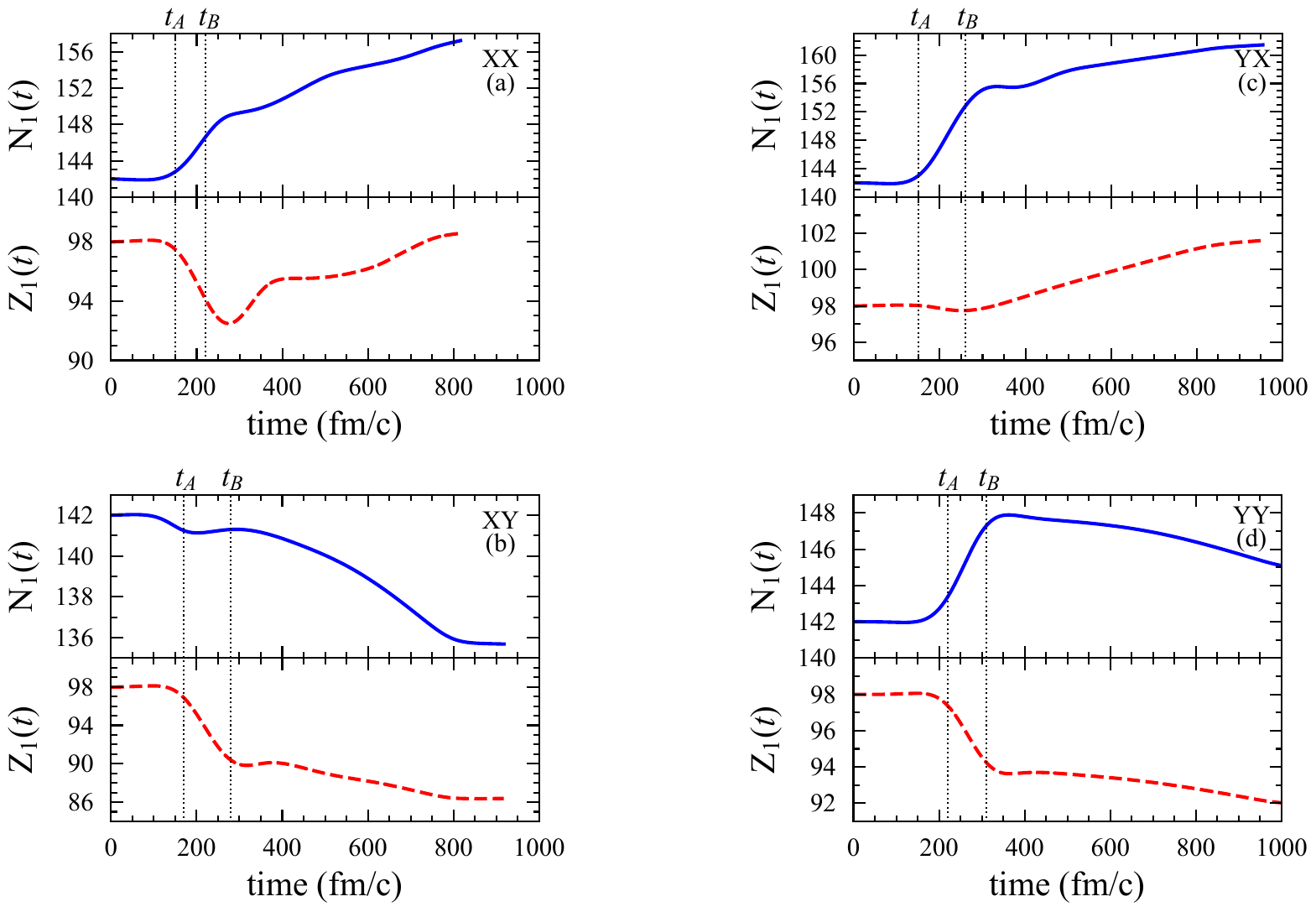}
\caption{Mean-values of neutron and proton numbers of Cf-like fragments in the
head-on collision of ${}^{240}\text{Cf}+{}^{246}\text{Th}$ system at
$E_\text{c.m.}=950$~MeV  in tip-tip (XX), tip-side (XY), side-tip (YX) and side-side
(YY) geometries. }
\label{fig2A}
\end{figure*}

\begin{figure*}[!th]
\includegraphics*[width=14cm]{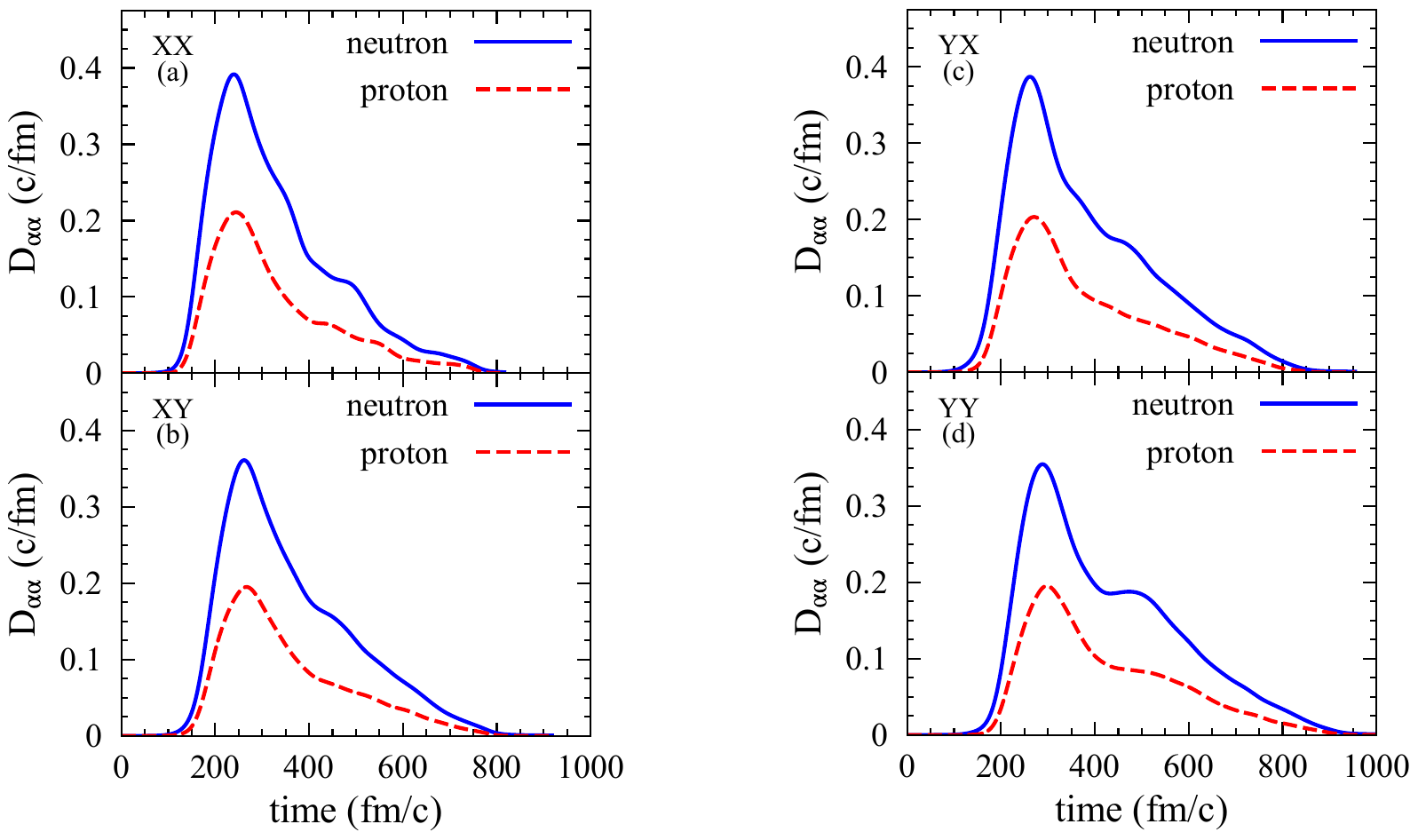}
\caption{Diffusion coefficients in the head-on collision of
${}^{240}\text{Cf}+{}^{246}\text{Th}$ system at $E_\text{c.m.}=950$~MeV  in tip-tip
(XX), tip-side (XY), side-tip (YX) and side-side (YY) geometries. }
\label{fig3A}
\end{figure*}
\bibliography{VU_bibtex_master}

\end{document}